\documentclass[preprint]{aastex}

\begin{document}
\title{Tentative Detection of Galaxy Spin Correlations in the Tully Catalogue}
\author{Ue-Li Pen}
\affil{Canadian Institute for Theoretical Astrophysics,
Toronto, Ont. M5S 3H8, Canada}
\email{pen@cita.utoronto.ca}
\author{Jounghun Lee}
\affil{Institute for Astronomy and Astrophysics, Academia Sinica,
Taipei, Taiwan}
\email{taiji@asiaa.sinica.edu.tw}
\author{Uro\v s Seljak}
\affil{Physics Department, Princeton University, Princeton, 
NJ 08544, USA}
\email{uros@feynman.princeton.edu}

\newcommand{\etal}{{\it et al.}}
\newcommand{\beq}{\begin{equation}}
\newcommand{\eeq}{\end{equation}}
\newcommand{\ben}{\begin{eqnarray}}
\newcommand{\een}{\end{eqnarray}}
\newcommand{\hbL}{\hat{\bf L}}
\newcommand{\hbS}{\hat{\bf S}}
\newcommand{\bS}{{\bf S}}
\newcommand{\bL}{{\bf L}}
\newcommand{\hL}{\hat{L}}
\newcommand{\hT}{\hat{T}}
\newcommand{\bk}{{\bf k}}
\newcommand{\by}{{\bf y}}
\newcommand{\hby}{\hat{\bf y}}
\newcommand{\bx}{{\bf x}}
\newcommand{\hbx}{\hat{\bf x}}
\newcommand{\hbr}{\hat{\bf r}}
\newcommand{\br}{{\bf r}}
\newcommand{\bI}{{\bf I}}
\newcommand{\bT}{{\bf T}}
\newcommand{\tT}{\tilde{T}}
\newcommand{\tbT}{\tilde{\bf T}}
\newcommand{\hbT}{\hat{\bf T}}
\newcommand{\pr}{\prime}

\begin{abstract}

We report a tentative detection of spin correlations in the Tully 
catalogue of nearby galaxies.  We define a simple but nontrivial spin
correlation function, and find an analytic estimate of it in the frame
of the linear perturbation theory.  Then, we present the observed spin
correlation signal from the Tully galaxies with error bars.  The three
dimensional spin correlation turns out to be significant at the $97\%$ 
confidence level, detected out to a few $h^{-1}$ Mpc.  
This observed correlation is consistent with the theoretical prediction 
based on the gravitational instability picture of galaxy formation.    
An analysis of systematic errors is also presented.  
The observed strength of correlation may be sufficient to
significantly affect blank field of weak lensing searches.

\end{abstract}

\section{INTRODUCTION}

Considerable effort has been devoted to the observational detection of  
any preferred direction in the orientation of galaxy spin vectors 
in the last century.  
Although several observational studies with various sample sizes 
and methods have reported so far the existence of some pattern 
of galaxy alignments \citep{str-str78,bin82,fli88,kas-oka92}, 
there is no established evidence for significant 
preferential galaxy alignments yet \citep{dek85,hof-etal89,
han-etal95,cab-dic99}.  However, it has been recently 
pointed out that the sample size of galaxies in the past 
efforts were too small to detect weak galaxy alignments 
unambiguously \cite{cab-dic99}.  

Theoretically, galaxy spin alignments is also an interesting issue 
since its existence has been predicted by the standard theory 
addressing the origin of galaxy spins.    Very recently, 
intrinsic correlation of galaxy alignments has attracted 
sharp considerable attention mainly since it could  
contribute as a significant contaminant in blank fields 
of the statistical search for weak gravitational lensing 
\cite{cat-etal00, hea-etal00,cro-met00}.  
In addition, detection of the intrinsic galaxy alignments by 
itself could have a fundamental impact on cosmology since 
it might shed a light on the reconstruction of the initial 
density field through the galaxy spin field    
(Lee \& Pen 2000a, hereafter LP00a).  
  
This Letter is aimed for two goals: to estimate the strength of 
intrinsic galaxy spin correlations from a physical theory and to 
measure the correlation signal directly from observed galaxies.

\section{PHYSICAL ANALYSIS}

In the gravitational instability picture of galaxy formation, the angular 
momentum of a galaxy originates from the local shear due to the 
proto-galactic tidal interaction with the surrounding matter 
\cite{hoy49,pee69}.  
Using the 1st-order linear theory \citep{whi84, cat-the96} 
(its prediction on the galaxy spin axis has been shown to be 
quite valid even in nonlinear regime by LP00a),  LP00a have found a 
general quadratic expression for the expected unit galaxy spins  
given the intrinsic local shears:   
$\langle \hL_i \hL_j |\hbT \rangle  =
(1+a)\delta_{ij}/3  - a\hT_{ik} \hT_{kj}$ 
where $\hbL$ is a unit galaxy spin, $\hbT$ is a unit 
traceless shear tensor, and $a$ is a correlation parameter introduced 
by LP00a to measure the strength of the correlation between 
$\hbL$ and $\hbT$. 
If galaxy spins are random, $a$ is zero.  
While if spins are maximally correlated with the shears 
\cite{cat-the96}, $a$ is $3/5$ 
(it was mistakenly cited as unity in LP00a, 
see Appendix A in Lee \& Pen 2000b, hereafter LP00b).   
The real value of $a$ must be determined empirically. 
LP00a found $a = 0.24$ in their numerical simulations. 
 
The above expression of $\hL_i \hL_j$ given $\hbT$ with nonzero 
value of $a$ implies the existence of the spin-spin correlation 
due to the spatial correlation of the local shears. 
Here we define a 2nd order spin-spin correlation function by 
$\eta(r)\equiv\langle |\hbL(\bx)\cdot\hbL(\bx +\br)|^{2}\rangle 
-\eta_{0}$  where $\eta_0$ is the value of 
$\langle |\hbL(\bx_1)\cdot\hbL(\bx_2)|^{2}\rangle$
for the case of random spins. $\eta_{0} = 1/3$ for the three 
dimensional (3D) spins while $\eta_{0} = 1/2$ for the 2D. 
Note that $|\hbL(\bx_1)\cdot\hbL(\bx_2)|^{2}$ is just $\cos^2(\Delta\theta)$
where $\Delta\theta$ is the angle between the two unit spin vectors, and
$\eta_0$ is the expectation value of $\cos^2(\theta)$ for a random angle 
$\theta$.  Using only directional information, the correlation in angle
is most readily measured this way.  Since we do not measure the chirality
of each galaxy's rotation axis, the statistic must be symmetric
under 180 degree rotation.

Using $\hL_i \hL_j = (1+a)\delta_{ij}/3  - a\hT_{ik} \hT_{kj}$, 
one can estimate 
$\langle |\hbL(\bx_1)\cdot\hbL(\bx_2)|^{2}\rangle =    
\langle\hL_{i}\hL_{i}^{\pr}\hL_{j}\hL_{j}^{\pr}\rangle$ such that 
$\langle\hL_{i}\hL_{i}^{\pr}\hL_{j}\hL_{j}^{\pr}\rangle 
=\langle\hL_{i}\hL_{j}\hL_{i}^{\pr}\hL_{j}^{\pr}\rangle  
= \langle (1+a)\delta_{ij}/3 - a\hT_{ik} \hT_{kj}\rangle 
\langle (1+a)\delta_{ij}/3 - a\hT_{il}^{\pr}\hT_{lj}^{\pr}\rangle  
= 1/3 - a^2/3 + a^{2}
\langle\hT_{ik}\hT_{kj}\hT_{il}^{\pr}\hT_{lj}^{\pr}\rangle$.  
To calculate $\langle\hT_{ik}\hT_{kj}\hT_{il}^{\pr}\hT_{lj}^{\pr}\rangle$, 
we approximate the unit traceless shear tensor, $\hbT$ as a Gaussian 
variable of the traceless shear,  and apply the 
Wick theorem to derive  
$\langle\hT_{ik}\hT_{kj}\hT_{il}^{\pr}\hT_{lj}^{\pr}\rangle 
\approx 
\langle\hT_{ik}\hT_{kj}\rangle
\langle\hT_{il}^{\pr}\hT_{lj}^{\pr}\rangle + 
\langle\hT_{ik}\hT_{il}^{\pr}\rangle
\langle\hT_{kj}\hT_{lj}^{\pr}\rangle + 
\langle\hT_{ik}\hT_{lj}^{\pr}\rangle 
\langle\hT_{kj}\hT_{il}^{\pr}\rangle = 1/3 + \xi_{R}^{2}(r)/6$.  
Here $\xi_R(r)$ is the correlation of the 
top-hat convolved density field normalized by $\xi_{R}(0) = 1$. 

Hence, we finally have  
\beq
\eta(r) \approx A\xi_{R}^{2}(r) 
\label{eqn:scor}
\eeq  
with $A = a^{2}/6$ for the 3D case. 
Note that this theoretical prediction has no free parameter 
and follows directly from first principles and an N-body 
normalization of the shear-spin correlation amplitude of $a$. 
The approximation is made only when we apply the Wick Theorem. 
In fact, we have verified in numerical realization of a Gaussian 
random field that the Wick expansion on $\hT$ approximates the 
true correlation within a few percent relative error. 
For the case of 2D,  it can be shown with the same manner
that $A = 25a^{2}/96$ .  
For a detailed derivation of eq. [1], see Appendix H in LP00b.  

On galaxy scales, the power spectrum is well approximated 
by a power-law, $P(k) = k^{n}$ with $n=-2$, for which 
$\xi_{R}(r) \sim r^{-1}$.  Thus, eq.[\ref{eqn:scor}] predicts 
that for neighboring galaxies (a few $h^{-1}$ Mpc), the spin-spin 
correlation is of order 1\%, decreasing rapidly as $r^{-2}$. 
It will decrease more rapidly as one moves to larger scales.  
Thus, we expect negligible spin correlation signal at 
separations of more than a few $h^{-1}$ Mpc.

\section{OBSERVED SIGNAL}

For this study, we use a subsample of $12122$ spiral galaxies 
from the Tully catalogue (Tully 2000). The spiral galaxies are 
selected here as galaxies having a type 0-9 as listed in Third 
Reference Catalogue of Bright Galaxies 
(de Vaucouleurs \etal\ 1991, RC3). 
It is a all-sky low-redshift sample with median redshift 
of $5180$ km/s.  
Among data, we retain necessary five pieces of information for each 
spiral galaxy: the position angle ($\alpha$), axial ratio ($\cal R$), 
the equatorial coordinates (1950 equinox for RA and DEC),  
radial distance (redshift), and  absolute diameter. 
The information of the position angle and axial ratio 
comes from the Uppsala General Catalogue (UGC) in 
the northern hemisphere \cite{nil74} and the ESO-UGC Catalogue  
in the southern hemisphere \cite{lau82}.
 
In a local spherical coordinate frame (the radial direction is given 
as the line of sight of a galaxy), the spin vector of a spiral galaxy 
can be determined through $\alpha$ and $\cal R$ by  
$|\hL_r| = {\cal R}$, $\hL_{\theta}/\hL_{\phi} = \tan(\pi -\alpha)$ 
upto the fourfold degeneracy: $\hat{L}_r \rightarrow -\hat{L}_r$ and 
$\{\hL_{\theta}, \hL_{\phi}\} \rightarrow -\{\hL_{\theta}, \hL_{\phi}\}$. 
Since we deal with the spatial orientation of the spin axis,  
only the twofold degeneracy in $\hat{L}_r$
is relevant to us.   
The spin vector expressed in this local spherical coordinates 
can be easily transformed into the global Cartesian coordinates 
using the given galaxy positions. 
 
But before determining the spin axis of each spiral galaxy,  
we consider several possible sources of systematic errors which might 
be involved in measurements of the given $\alpha$ and $\cal R$.    
For example, deviation of the shape of spiral galaxies from perfect 
ellipses,  finite thickness of galaxies,  differences between 
the UGC and the ESO-UGC catalogue, and others. 
  
We redistribute the given $\alpha$ and $\cal R$ uniformly 
according to their original order ($\cal R$ is redistributed 
within each galaxy Hubble type).  We then have $\alpha$ perfectly
uniformly distributed in the interval $[0,\pi)$ and $\cal R$ in
$[0,1)$.  The redistribution of $\cal R$ statistically removes
finite disk thickness effects if each disk has the same thickness
within each Hubble type.  This uniform redistribution
reduces the final spin-spin correlation strength very slightly, 
indicating that we do not expect systematic effects of global 
biases to affect our results.  However, 
we would expect the shape-shape correlation of galaxies to result
in potential problems ($\cal R$-related systematic errors).  
So we choose to measure a statistic which does not depend on 
such correlations.

One practical way to calculate the true correlation signal  
free of $\cal R$-related systematic errors is to use the projected 
spins:  
Instead of using the given values of $\cal R$,  
we can set ${\cal R} = 0$ for all galaxies,
which amounts to projecting the 3D spin vectors 
onto the plane of the sky (the plane normal to the line of sight).
Using these projected 2D spins,   
one can construct a real (but 2D) spin correlation function, 
$\eta_{2D}$ as a function of the 3D galaxy separations as follows:
Let us consider two galaxies with unit spins $\hbL_{i}$ and 
$\hbL_{j}$ at $\hbx_{i}$ and $\hbx_{j}$ respectively with 
a separation $\br =\hbx_{i}-\hbx_{j}$, $\hbr = \br/|\br|$.  
The 2D projected unit spin $\hbS$ for each galaxy 
can be expressed as $\hbS_{i} = \bS_{i}/|\bS_{i}|$, 
$\bS_{i} = \hbL_{i} - (\hbL_{i} \cdot \hbx_{i})\hbx_{i}$, 
and the 2D projected unit separation vector is, 
$\hby_i= \by_i/|\by_i|$,  
$\by_{i}=\hbr -(\hbr\cdot\hbx_{i})\hbx_{i}$.  
The 2D spin correlations, $\eta_{2D}$ now can be expressed as the 
difference in angle, $\beta$, between $\hbS$ and $\hby$ such that   
$\eta_{2D} = \langle\cos^{2}(\beta_{i} - \beta_{j})\rangle - 1/2$ 
with $\cos\beta_{i} = \hbS_{i}\cdot\hby_{i}$. This procedure 
reduces to $|\hbS_{i}\cdot \hbS_{j}|^2-1/2$ in the small angle 
(flat sky) limit, but is also valid for all-sky data used here.     
Note that the two-fold degeneracy, ($\hL_r \longrightarrow -\hL_r$) 
is irrelevant for the 2D case where $\hL_r = 0$. 

To find a 3D spin correlation signal free of the systematic 
errors involved in $\cal R$,  one can consider the correlation  
between the 3D and 2D unit spin vectors such that 
$\eta(r) = \langle |\hbL\cdot\hbS|^2\rangle - 1/3$ 
where $\hbS$ and $\hbL$ are calculated from 
the uniformly redistributed $\alpha$ and $\cal R$.  
Here $\hbS$ is free of the $\cal R$-related 
systematic errors by setting $\cal R$ = 0, 
but $\hbL$ still involves $\cal R$-related systematic errors.  
Thus, some fraction of $\eta(r)$ must be the false signal. 
As a device from which the false signal is to be filtered out, 
one can use the following procedure: 
Generate random $12122$ $\alpha$'s.  Using these random $\alpha$'s, 
construct 2D random spins, $\hbS^{\prime}$.   
Then calculate the correlation of $\eta^{\pr}(r) \equiv 
\langle |\hbL\cdot\hbS^{\prime}|^2\rangle - 1/3$.   
Repeat this process 1,000 times with different sets of 12122 
random $\alpha$'s, and take the average, $\bar{\eta}^{\pr}(r)$ 
out of the 1000 sets of the correlations.   Then $\bar{\eta}^{\pr}(r)$ 
represents purely the false signal only, otherwise $\bar{\eta}^{\pr}(r)$ 
would be almost zero.  The effective 3D correlation free of 
the $\cal R$-related systematic errors is now obtained  
as $\eta_{E}(r) = \eta(r) - \bar{\eta}^{\pr}(r)$. 

One thing one has to keep in mind for the 3D case, however, 
is the twofold degeneracy.  As explained in $\S 2$, one cannot 
determine the sign of the radial component of a galaxy  
spin.   Let us assume that we have a galaxy pair whose 
3D unit spins and projected 2D unit spins are 
$\hbL_{i}$, $\hbS_{i}$ and $\hbL_{j}$, $\hbS_{j}$ respectively.
Here each 3D unit spin is determined upto twofold ambiguity, say  
$\{\hbL^{a}_{i}$,   $\hbL^{b}_{i}\}$, 
$\{\hbL^{a}_{j}$,   $\hbL^{b}_{j}\}$ ($\hbL^{a}$ and $\hbL^{b}$ differ 
by sign of their radial components).
Thus, there are four different ways to correlate 
the spins of this galaxy pair such that 
$f^{a}_{ij} \equiv |\hbL^{a}_{i}\cdot\hbS_{j}|^2$, 
$f^{b}_{ij} \equiv |\hbL^{b}_{i}\cdot\hbS_{j}|^2$, 
$f^{a}_{ji} \equiv |\hbL^{a}_{j}\cdot\hbS_{i}|^2$, 
$f^{b}_{ji} \equiv |\hbL^{b}_{j}\cdot\hbS_{i}|^2$. 
The spin correlation of this galaxy pair given the twofold 
degeneracy is now taken as the average of the four combinations:  
$(f^{a}_{ij}+f^{b}_{ij}+f^{a}_{ji}+f^{b}_{ji})/4 -1/3$. 
We apply this to every galaxy pair in calculating $\eta(r)$. 

\begin{figure}
\plotone{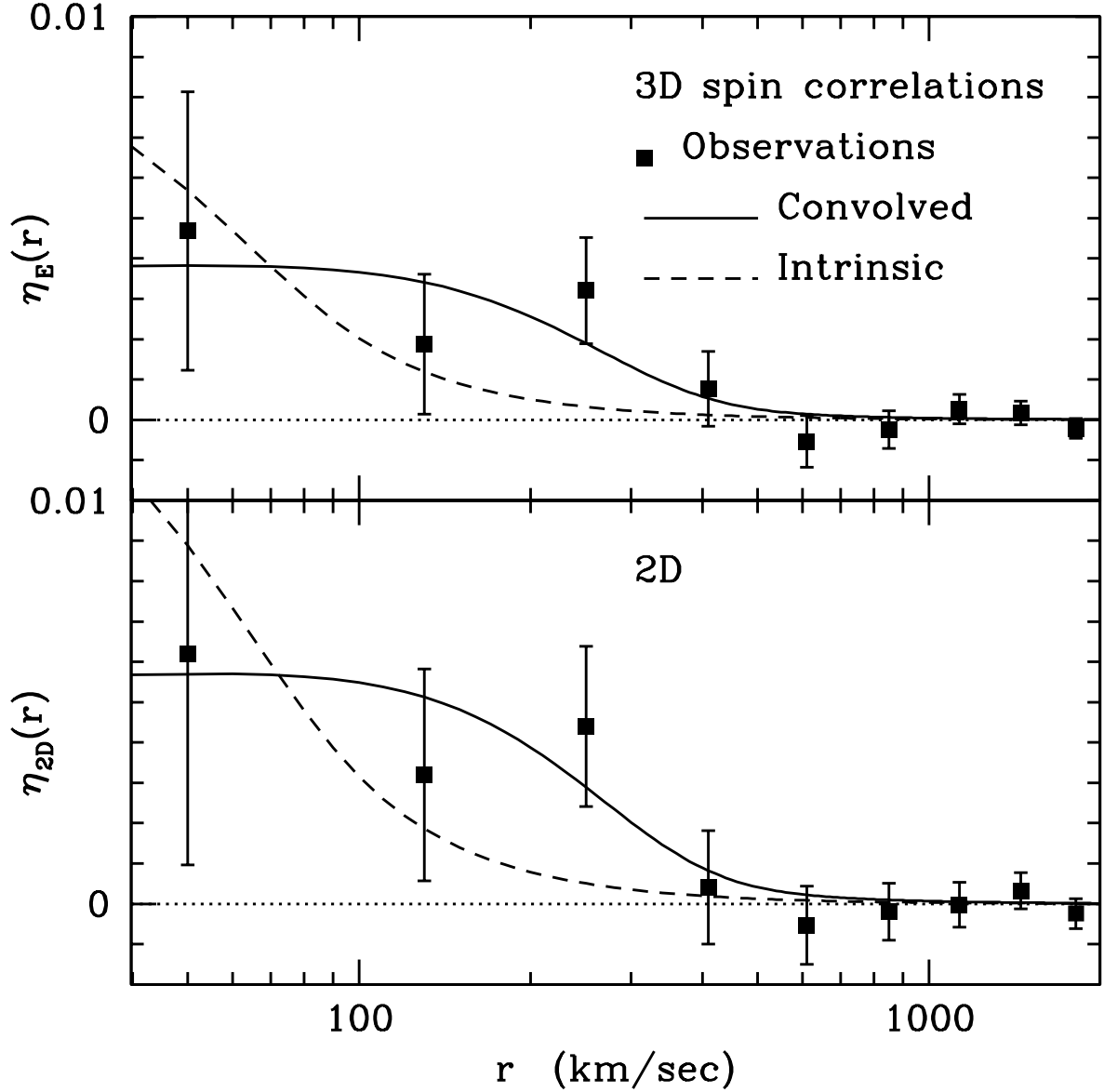}
\caption[spinfig1.ps]
{The spin-spin correlation as a function of galaxy 
separations, $r = cz$ (km/s). The power-law spectrum, $P(k) = k^{-2}$ 
and the correlation parameter, $a=0.24$ are used for the theoretical 
curves. The solid theoretical curve is a Gaussian convolved version of the
dashed one with a peculiar-velocity dispersion of $\sigma_{v} = 150$ km/s.} 
\end{figure}

Fig.1 plots the resulting $\eta_{2D}$ and $\eta_{E}(r)$ (the square dots) 
with error bars, and compares it with the theoretical predictions 
(eq. [\ref{eqn:scor}]).  
The error bars are from the numerical formula 
which we have obtained by the Monte-Carlo method.  We have generated 
$N_b$ samples of $10^{6}$ random spins, and calculate the standard 
deviations of the correlations of the random spins 
We have found that the standard deviations are excellently 
fitted by the following formula: 
$0.234/\sqrt{N_{b}}$ for the 3D, and  $1/\sqrt{8N_{b}}$ 
for the 2D.  For the Tully galaxies, $N_b$ corresponds to 
the number of galaxy pairs in each bin.  
  
Note that the observed correlation is measured in redshift space, 
which causes the observed signal to look flatter than the intrinsic 
Lagrangian  correlation function defined in real space (the dashed line).   
In order to make a better comparison, we convolve the Lagrangian 
correlation by a 1D Gaussian filter with a peculiar-velocity dispersion 
of $\sigma_{v} = 150$ km/s to obtain the theoretical 
curve in redshift space (the solid line).  Since we restrict ourselves
to spiral galaxies, we use a value of $\sigma_{v}$ close to the 
observed dispersion of $126 \pm 10$ km/sec \cite{dav-etal97}
for this class of galaxies.  
We note that this very cold distribution of field spirals implies
that galaxies did not move far from their initial positions, leaving
nearest neighbor relations relatively unchanged from their Lagrangian
values.

The 1D Gaussian filter instead of a 3D one is used to account for 
the non-uniform galaxy distribution. Since galaxies are not uniformly 
distributed, one has to weight $\eta(r)$ by the galaxy-galaxy 
correlation which is proportional $r^{-1.8}$ when convolving it.  
This effectively cancels out the 3D volume factor of $r^2$.  
For the theoretical curves, a power-law spectrum of $P(k) = k^{-2}$  
and the correlation parameter of $a = 0.24$ are used.  
The spin-spin correlation is of much smaller amplitude of  
$a^2 \xi(r)^2/\xi(0)^2$ than $a$, and thus is much harder to measure 
in simulations.  Croft \& Metzler (2000) and Heavens \etal\ (2000) 
have attempted it and found signals of similar amplitudes. 
But due to their different statistics from ours, it is difficult 
to convert their findings into ours.
  
The observed correlation signal is in fairly good accord with the 
theoretical predictions.  The total significances of the observed 
2D and the 3D correlations at the first three bins are at detectable 
$95\%$ and $97\%$ confidence levels respectively within a distance of 
a few $h^{-1}$Mpc, while the correlations diminish to zero at larger
separations as the theory predicts (see $\S 2$).  For the total
significance, we evaluate the $\chi^{2}$ from the first three bins
(degrees of freedom, $n_D = 3$) with the null hypothesis of random
spins.  Then, we calculate the probability of $\chi^{2}$ exceeding the
observed value ($\chi^{2}_{2D} = 7.78$, $\chi^{2}_{3D} = 8.95$) and
show that the null hypothesis of random spins, i.e., no correlation
hypothesis at the first three bins is rejected at $95\%$ and $97\%$
confidence levels for the 2D and the 3D cases respectively.  For this
$\chi^2$ distribution, the 2D and the 3D correlations are significant
at $2.0\sigma$ and $2.4\sigma$ level respectively.

Another concerns that could arise in measuring the spin 
correlation signal is the {\it overlap effect}. 
When a separation of a pair galaxy is less than the physical size 
of the galaxy pair,  then the spin correlation signal calculated 
from that close pair might be affected by the overlap effect.  
We remeasure the signal excluding those close pairs 
whose separations are less than twice the sum of the 
individual radii of the galaxy pair (total 279 galaxy pairs excluded). 
But it turns out that  the excision to twice the radius at which 
overlap is expected has a negligible effect on the signal 
(the total significance of the observed signal is slightly 
lowered from $97\%$ to $95\%$ for the 3D, from $95\%$ to $94\%$ 
for the 2D). 

It is also interesting to measure the covariance of $\eta(r)$, 
${\rm cov}[\eta(r_i),\eta(r_j)] \equiv 
\langle\eta(r_i)\eta(r_j)\rangle/(\sigma_{\eta_i}\sigma_{\eta_j})$ 
between the bins using the 1,000 sample sets of random spins. 
It provides us the degree of independence of the statistics 
at each bin.  We have found 
${\rm cov}[\eta(r_i),\eta(r_j)] \approx {\cal O}(10^{-3}) \ll 1$ 
for both the 2D and 3D cases, suggesting the statistics at each bin 
is quite independent of that of the neighboring bins.

\section{DISCUSSIONS AND CONCLUSIONS}

There are other sources of shear which could change the apparent
direction of the spin axis.  Above all, shears from weak lensing can
be a good example.  But the shear effect on the galaxy-spin axis from
the weak lensing is much smaller in this low redshift sample: To
estimate the magnitude of the weak lensing shear, one has to consider
the spatial correlation of the {\it convergence}, $\kappa$ defined as
the ratio of the observed galaxy surface density to the critical
surface density (the surface density which can just refocus the light
beam in the case of null shear).  But the expected weak lensing shear
is less than a percent at $1^{\circ}$
for sources at redshift $z \sim 1$ and grows as $\kappa
\propto z^{0.6}$ \cite{JS97}, so the expected
weak lensing shear effect on the galaxy spin axis for our sample, $z
\approx 0.02$ is ${\cal O}(10^{-3})$ at  
a separation of 1 $h^{-1}$Mpc,
one order of magnitude smaller than that of the local cosmic shear we
measure.

Given that the dominant shear effect on the galaxy alignments 
at low redshift is from the local cosmic shear, 
the detection of galaxy alignments could affect 
recent observational surveys for the weak lensing effect on the distortion 
of the galaxy shapes \cite{wae-etal00} and the interpretation 
of the lensing effect \citep{cat-etal00,hea-etal00}.
It is beyond the scope of this letter to derive a detailed
quantitative calculation of the impact of the observed shape-shape
correlation on weak lensing surveys, which is being addressed in a
separate paper (Crittenden et al. 2000, in preparation).  
Here we will make a brief qualitative estimate which shows that the 
intrinsic alignments may introduce a non negligible  bias.

To estimate the impact on weak lensing, we project the 3D 
spin-spin correlation (eq. [\ref{eqn:scor}]) using the small angle
approximation onto the sky to obtain a projected 2D angle-angle
correlation weighted by the galaxy-galaxy correlation function 
[$\xi(r) \sim (r/r_0 )^{-1.8}$ with $r_{0} = 5.5 h^{-1}$ Mpc] : 
\ben
\eta_2(y)=\frac{\int_{-{\rm L}}^{{\rm L}}\eta_{2D}
(\sqrt{y^2+z^2}) [(1+\xi(y^2+z^2)] dz}{
\int_{-{\rm L}}^{\rm L} 1+\xi(y^2+z^2) dz}.
\label{eqn:eta2}
\een
L is the {\it depth} of the galaxy distribution in the source plane.  
The dilution effect causes $\eta_2 \propto$ 1/L, and the usual
approximation is to make ${\rm L} \longrightarrow \infty$, stating that 
two galaxies that appear close in projection are not actually close in
real space.  The fact that galaxies cluster to each other modifies
this assumption, such that galaxies that are close in projection
actually have an intrinsic probability of being neighbors that is
enhanced by the correlation function.  To estimate the effect, we use
L =1 Gpc (comoving) and model $\eta_{2D}=0.01/r$ 
for $r>1 $ Mpc and $\eta_{2D}=0.01$ at smaller separations.  
Placing the source at a typical
comoving angular diameter distance of $2800 h^{-1}$ Mpc (corresponding to
$z\sim 1$ in a $\Omega_0=.65$, $h=0.7$ universe, see Pen 1999) and
assuming non-evolving comoving correlation functions, we find an apparent
correlation at one arcminute of $\eta_2(1') \sim 0.00024$, dropping as
$1/y$ at larger scales.

Weak lensing surveys measure the correlation function of the shear 
$\gamma_{i}$ at separation $r$,
where $\gamma_i$, $i=1,2$, are the two components 
of the symmetric traceless shear tensor. 
Since equation (\ref{eqn:eta2}) correlates galaxies assuming unit
ellipticities, it gives a typical intrinsic correlation induced shear 
amplitude $c(r)=\langle\gamma_1(0)\gamma_1(r)+
\gamma_2(0)\gamma_2(r)\rangle=\langle e(0)e(r) 
\cos 2 \Delta \theta \rangle=2\langle e\rangle^{2}\langle 
\cos^2\Delta \theta-1/2 \rangle=
2\langle e \rangle^2\eta_{2D}(r) \sim 2 \times 10^{-4}$, where 
we used $\langle e^2\rangle^{1/2}=0.6$ as given by Kaiser (1998). 
We note that recent detection on one arcminute scale 
(van Waerbeke et al. 2000) finds a candidate weak lensing
signal for a Top-Hat convolved shear variance, 
which  gives $c(1')\sim 5 \times 10^{-4}$ 
when translated into our correlation function. 
This is somewhat larger, but still of comparable
magnitude and angular dependence as our intrinsic alignment estimate.
This model has made many simplifying assumptions, most crucial one of
which is that the shape-shape and galaxy-galaxy correlation functions
do not evolve, and that the depth distribution of source galaxies is
$1$ Gpc, all of which may be uncertain by at least a factor of 2.  A
direct measurement of each of these quantities is essential to fully
quantify the impact of shape-shape alignment on weak lensing.  We
note, however, that one test can be made which would probe for the
presence of intrinsic correlations.  The observed shear field
$\gamma_{ij}$ can be decomposed into an ``electric'' ($E$), or
potential flow component, and a ``magnetic'' ($B$) or curl component
\cite{pen00}.  
Weak lensing predicts $B=0$, while this is not in
general the case for shape-shape alignment, which predicts a comparable
strength of $E$ and $B$ signal.  A measurement of $B$
would thus provide a general control experiment.

We have estimated the expected strength of the spin correlation 
normalized to numerical simulation results (LP00a) in the frame 
of the 1st order linear perturbation theory.  Then we have measured  
the spin correlation signal directly from the observed spiral galaxies 
from the Tully catalogue, and tested our theoretical predictions 
against the observed signal.  Both the predicted and observed 
amplitudes are of order of $1 \%$ at a separation of $1$ 
$h^{-1}$Mpc, in agreement with each other.  
Given our results, we conclude that past surveys would not have had 
enough sensitivity to detect this signal.  

Qualitative estimates show that these correlations may have a nonnegligible
effect on statistical weak lensing surveys.  We have proposed using
the ``magnetic'', or curl, component of the apparent shape-shape
alignment to discriminate between intrinsic shear and lensing induced
shear. 

\acknowledgements 

We are very grateful to B. Tully for his catalogue and helpful 
discussions. We also thank the referee who helped us improve the 
manuscript.  U. Pen thanks L. Van Waerbeke for useful comments.  
This work has been supported by Academia Sinica and partially by  
NSERC grant 72013704 and computational resources of the 
National Center for Supercomputing Applications. U. Seljak  
acknowledges the support of NASA grant NAG5-8084.


\begin{thebibliography}{}
\bibitem[Binggeli 1982 ]{bin82}
Binggeli, B. 1982, \aap, 107, 338 
\bibitem[Cabanela \& Dickey 1999]{cab-dic99}
Cabanela, J. E., \&  Dickey, J. M. 1999, \aj, 118, 46  
\bibitem[Catelan \& Theuns 1996]{cat-the96} 
Catelan, P., \& Theuns, T. 1996, \mnras, 282, 436 
\bibitem[Catelan \etal\ 2000]{cat-etal00} 
Catelan, P., Kamionkowski, M., \& Blandford, R. 2000, preprint 
(astro-ph/0005470) 
\bibitem[Croft \& Metzler 2000]{cro-met00}
Croft, R. A. C., \& Metzler, C. A. 2000, preprint (astro-ph/0005384)
\bibitem[Davis \etal\ 1997]{dav-etal97}
Davis, M., Miller, A., \& White, S. 1997, \apj, 490, 63 
\bibitem[Dekel 1985]{dek85} 
Dekel, A. 1985, \apj, 298, 461 
\bibitem[de Vaucouleurs \etal\ 1991]{dev_etal91}
de Vaucouleurs, G., de Vaucouleurs, A.,  Corwin, H. G., Jr., Buta, R. J., 
Patuerl, G., \& Fouque, P. 1991, Third Referece Catalogue of Bright 
Galaxies (New York: Springer) (RC3) 
\bibitem[Doroshkevich 1970]{dor70} 
Doroshkevich, A. G. 1970, Astrofisika, 6, 581
\bibitem[Flin 1988]{fli88}
Flin, P. 1988, \mnras, 235, 857 
\bibitem[Han \etal\ 1995]{han-etal95} 
Han, C., Gould, A., \&  Sacket, P. D. 1995, \apj, 445, 46 
\bibitem[Heavens \etal\ 2000]{hea-etal00} 
Heavens, A., Refregier, A., \& Heymans, C. 2000, preprint 
(astro-ph/0005269) 
\bibitem[Helou \& Salpeter 1982]{hel-sal82}
Helou, G. , \& Salpeter, E. 1982, \apj, 252, 75  
\bibitem[Hoyle 1949]{hoy49} Hoyle, F. 1949, in Problems of 
Cosmical Aerodynamics, ed. J. M. Burgers \& H. C. van de Hulst 
(Dayton : Central Air Documents Office), 195
\bibitem[Hoffman \etal\ 1989]{hof-etal89}
Hoffman, G. L., Lewis, B. M., Helou, G., Salpeter, E. E., \& 
Williams, H. L. 1989, \apjs, 69, 95 
\bibitem[Jain \& Seljak 1997]{JS97}
Jain, B., \& Seljak, U. 1997, ApJ, 484, 560
\bibitem[Kaiser 1998]{Kaiser98}
Kaiser, N. 1998, ApJ, 498, 26
\bibitem[Kashikawa \& Okamura 1992]{kas-oka92}
Kashikawa, N., \& Okamura, S. 1992, \pasj, 44, 493  
\bibitem[Lauberts 1982]{lau82}
Lauberts, A. 1982, The ESO/Uppsala Survey of the ESO (B) Atlas 
(Garching: ESO) 
\bibitem[Lee \& Pen 2000a]{lp00a} 
Lee, J., \& Pen, U. 2000, \apj, 532, L5 (LP00a)
\bibitem[Lee \& Pen 2000b]{lp00b}
Lee, J., \& Pen, U. 2000, preprint (astro-ph/0008135) (LP00b)
\bibitem[Nilson 1974]{nil74}
Nilson, P. 1974, Uppsala Astron. Obs. Annals, 6 
\bibitem[Peebles  1969]{pee69}
Peebles, P. J. E., 1969, \apj, 155, 393 
\bibitem[Pen 1999]{pen99}
Pen, U. 1999, \apjs, 120, 49
\bibitem[Pen 2000]{pen00}
Pen, U. 2000, \apj, 534, L19
\bibitem[Strom \& Strom 1978]{str-str78}
Strom, S. E., \& Strom, K. M. 1978, \aj, 83, 732 
\bibitem[Tully 2000]{tul00}
Tully, B. 2000, private communication 
\bibitem[Van Waerbeke \etal\ 2000]{wae-etal00} 
Van Waerbeke \etal\ 2000,  \aap, in press 
\bibitem[White 1984]{whi84} 
White, S. D. M. 1984, \apj, 286, 38 
\end{thebibliography}
\end{document}